\documentclass[useAMS,usenatbib,usegraphicx]{mn2e}
\usepackage{aas_macros}
\usepackage{amsmath}
\usepackage{amssymb}
\usepackage{graphicx}
\usepackage{txfonts}

\title[Perseus I \& NGC 3109 association in the LG context]{Perseus I and the NGC 3109 association in the context of the Local Group dwarf galaxy structures}
\author[Pawlowski \& McGaugh]{Marcel S. Pawlowski\thanks{E-mail:
marcel.pawlowski@case.edu}, Stacy S. McGaugh \\
Department of Astronomy, Case Western Reserve University,
              10900 Euclid Avenue, Cleveland, OH, 44106}
\begin{document}
\date{Accepted 2014 February 13.  Received 2014 February 13; in original form 2014 January 8}
\pagerange{\pageref{firstpage}--\pageref{lastpage}} \pubyear{2014}
\maketitle
\label{firstpage}
\begin{abstract}
The recently discovered dwarf galaxy Perseus I appears to be associated with the dominant plane of non-satellite galaxies in the Local Group (LG). We predict its velocity dispersion and those of the other isolated dSphs Cetus and Tucana to be 6.5, 8.2, and $5.5\,\mathrm{km\,s}^{-1}$, respectively. The NGC 3109 association, including the recently discovered dwarf galaxy Leo P, aligns with the dwarf galaxy structures in the LG such that all known nearby non-satellite galaxies in the northern Galactic hemisphere lie in a common thin plane (rms height 53 kpc; diameter 1.2 Mpc). This plane has an orientation similar to the preferred orbital plane of the Milky Way (MW) satellites in the vast polar structure. Five of seven of these northern galaxies were identified as possible backsplash objects, even though only about one is expected from cosmological simulations. This may pose a problem, or instead the search for local backsplash galaxies might be identifying ancient tidal dwarf galaxies expelled in a past major galaxy encounter. The NGC 3109 association supports the notion that material preferentially falls towards the MW from the Galactic south and recedes towards the north, as if the MW were moving through a stream of dwarf galaxies.
\end{abstract}

\begin{keywords}
Galaxies: dwarf -- Galaxies: individual: Perseus I -- Galaxies: kinematics and dynamics -- Local Group -- dark matter
\end{keywords}

\section{Introduction}

The Milky Way (MW) is surrounded by a vast polar structure (VPOS) of satellite objects including the satellite galaxies, young halo globular clusters and several stellar and gaseous streams \citep*{1976MNRAS.174..695L,2012MNRAS.423.1109P}. The proper motions of the 11 classical satellite galaxies reveal that these almost exclusively co-orbit in this VPOS, which allowed us to predict the proper motions of the remaining satellite galaxies \citep{2013MNRAS.435.2116P}. 

\citet*{2005A&A...431..517K} have first identified this planar alignment as being inconsistent with cosmological simulations based on the cold dark matter paradigm with a cosmological constant, $\Lambda$CDM. This finding subsequently triggered an ongoing debate on whether such structures can be reconciled with cosmological expectations \citep[e.g.][]{2008ApJ...686L..61D,2008MNRAS.385.1365L,2009ApJ...697..269M,2009MNRAS.399..550L,2011MNRAS.415.2607D,2012MNRAS.424...80P,2013MNRAS.429.1502W,2013MNRAS.435.2116P} or rather points at a different origin such as the formation of phase-space correlated tidal dwarf galaxies \citep[TDGs, e.g.][]{2005PASJ...57..429S,2007MNRAS.376..387M,2010ApJ...725L..24Y,2011A&A...532A.118P,2012MNRAS.427.1769F,2013MNRAS.431.3543H,2013MNRAS.429.1858D}.

\citet{2013Natur.493...62I} and \citet{2013ApJ...766..120C} have recently discovered a similar 'Great Plane of Andromeda' (GPoA), a co-orbiting alignment consisting of about half of the satellite galaxies of the Andromeda galaxy (M31), the other major galaxy in the Local Group (LG). Motivated by this discovery that satellite galaxies appear to preferentially live in phase-space correlated structures, \citet{2013MNRAS.435.1928P} set out to search for similar structures on a LG scale. They have discovered that all but one of the 15 LG dwarf galaxies more distant than 300\,kpc from the MW and M31 are confined to two narrow (short-to-long axis ratios of $\approx 0.1$) and highly symmetric planes, termed LGP1 and LGP2. LGP1 is the dominant plane both by number of objects (about nine), and alignment with additional features, such as the Magellanic Stream which traces the positions and line-of-sight (LOS) velocities of the LGP1 plane members in the southern Galactic hemisphere.

Given that the number of known dwarf galaxies in the LG more distant than 300\,kpc from both major galaxies is still low, each additional detection poses a chance to test the existence of the planar LG structures and to potentially refine our understanding of these structures. Such an opportunity is now provided by the recent discovery of the dwarf spheroidal (dSph) galaxy Perseus I at a distance of 374\,kpc from M31 \citep{2013arXiv1310.4170M}. In the following, we test whether it can be considered to be associated to either LGP1 or LGP2.

In addition, we predict the velocity dispersion of Perseus I and two other non-satellite dSphs as expected in Modified Newtonian Dynamics \citep[MOND,][]{1983ApJ...270..365M,2012LRR....15...10F}. Similar predictions have been made for other dwarf galaxies in M31's vicinity \citep{2013ApJ...766...22M} and these have successfully passed the test of observations \citep{2013ApJ...775..139M}. Unfortunately, no similar predictions are possible in the $\Lambda$CDM framework.

Another recently discovered nearby dwarf galaxy, Leo P \citep{2013AJ....146...15G,2013AJ....145..149R}, has lead \citet{2013arXiv1310.6365B} to re-investigate the NGC 3109 association, a group of dwarf galaxies at a distance of about 1.3--1.4\,Mpc from the MW that consists of NGC 3109, Antlia, Sextans A and Sextans B \citep{1999ApJ...517L..97V,2006AJ....132..729T}. They realised that Leo P aligns with the four other members of the association in a very narrow, linear structure. As the NGC 3109 association is very close to the LG and has a linear extent of 1.2\,Mpc, similar to its distance from the MW, we will investigate its orientation in the context of the LG planes of non-satellite dwarf galaxies. This reveals an intriguing alignment with the other three nearby non-satellite galaxies in the northern hemisphere of the MW and leads us to discuss suggested origins for the NGC 3109 association in light of the geometry of the LG.

The paper is structured as follows. In Sect. \ref{sect:perseus} we determine whether Perseus I is associated with one of the dwarf galaxy planes in the LG. In Sect. \ref{sect:MOND} we predict the velocity dispersion of the distant dSphs in the LG, Perseus I, Cetus and Tucana. In Sect. \ref{sect:NGC3109association} we determine the orientation of the NGC 3109 association in the same coordinate system used in \citet{2013MNRAS.435.1928P}, discuss possible origins for the alignment and conclude that the association is likely part of the LG dwarf galaxy structures. In Sect. \ref{sect:backsplash} we discuss how the search for cosmological backsplash galaxies in the LG might give rise to two additional small-scale problems of cosmology and how it could falsely identify TDGs as backsplash objects. In Sect. \ref{sect:scheme} we present a sketch of the LG dwarf galaxy structures and their preferred direction of motion and discuss open questions and limitations in Sect. \ref{sect:openquestions}. Finally, we summarize our results in Sect. \ref{sect:conclusion}.

\section{Perseus I and the LG planes}
\label{sect:perseus}

\begin{figure}
   \centering
   \includegraphics[width=85mm]{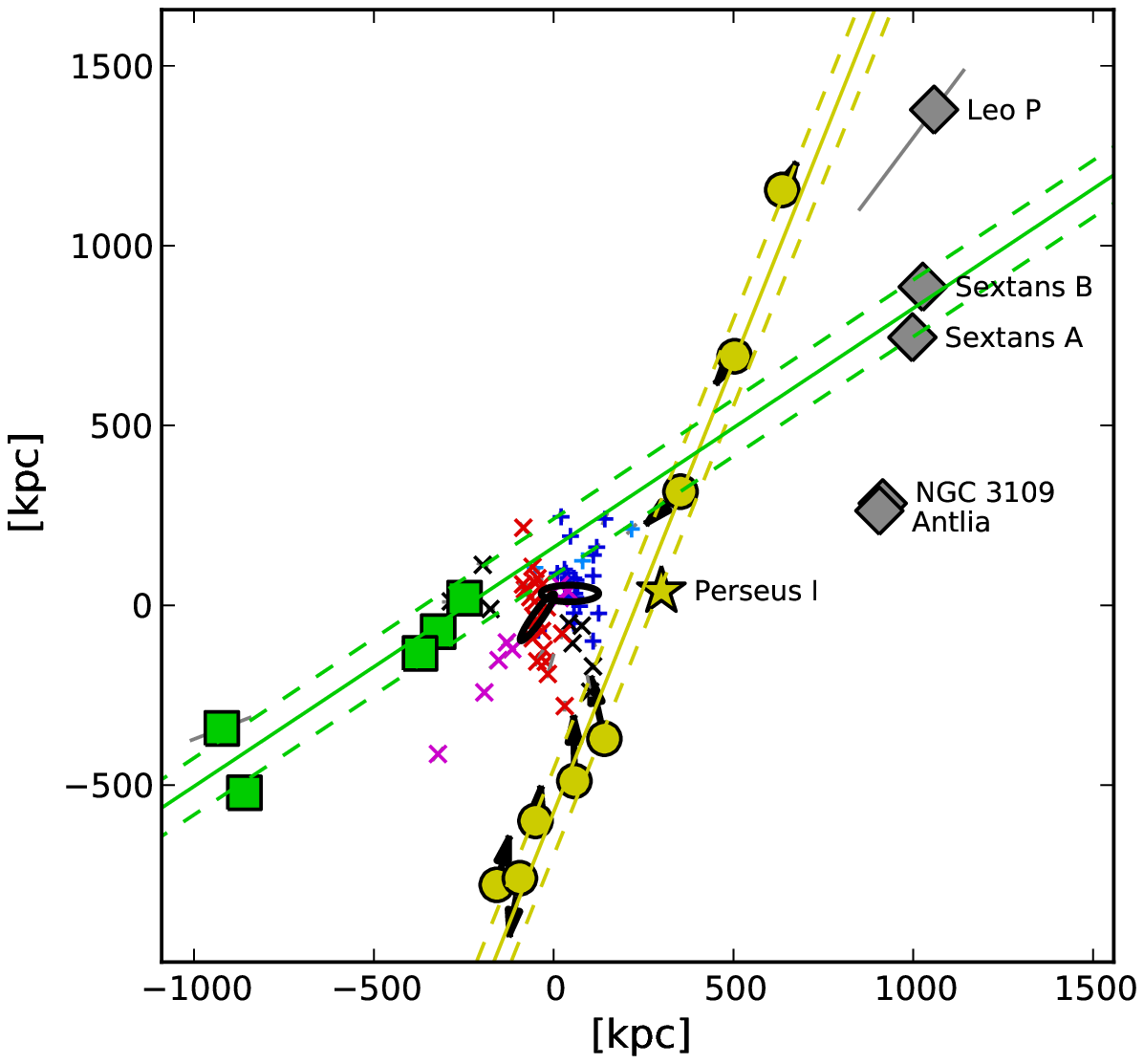}
   \includegraphics[width=85mm]{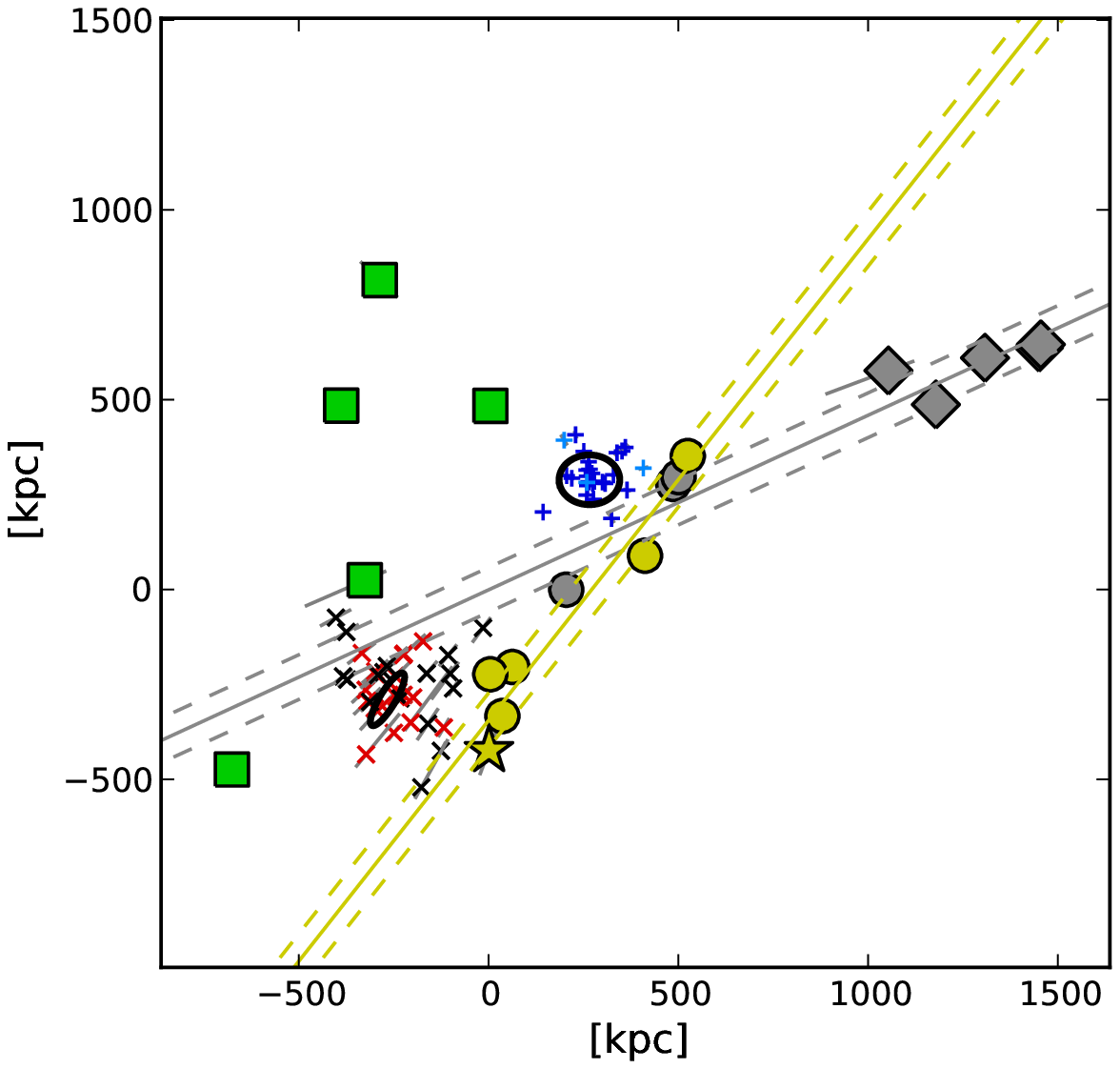}
   \caption{\textit{Top panel}: edge-on view of the non-satellite dwarf galaxy planes LGP1$^{\mathrm{mod}}$\ (yellow dots) and LGP2 (green squares) in the LG (looking along $[l, b] = [308.9^\circ, 16.8^\circ]$). The best-fit planes are plotted as solid lines, the dashed lines denote their RMS heights. The black ellipses indicate the positions and orientations of the MW and M31 and the blue plus signs (red and black crosses) indicate MW (M31) satellites. Grey lines mark the 1$\sigma$\ distance uncertainties.
   The newly discovered dwarf galaxy Perseus I (yellow star) is aligned with LGP1$^{\mathrm{mod}}$. Compare to the similar plot shown as Fig. 9 in \citet{2013MNRAS.435.1928P}. 
   In addition, the plot shows the positions of the dwarf galaxies associated to the linear NGC 3109 association (grey diamonds) which is situated behind the LG in this view. The NGC 3109 association is almost parallel to the dominant non-satellite plane in the Local Group, but offset by 0.3--0.5\,Mpc.
   \textit{Bottom panel}: Edge-on view of LGP1$^{\mathrm{mod}}$\ and the plane fitted to the five members of the NGC 3109 association plus the LGP1$^{\mathrm{mod}}$\ members Leo T, Leo A and UGC 4879 (looking along $[l, b] = [348.3^\circ, -54.9^\circ]$; the view of the upper panel would originate from approximately the lower left of this plot). The NGC 3109 association might be related to the dominant LG plane (LGP1$^{\mathrm{mod}}$). The planes are inclined to it by only $27^\circ$ and the intersection of the two planes lies close to the MW.
   }
              \label{fig:LGplanes}
    \end{figure}
    

\begin{table}
 \small
 \caption{Parameters of the fitted planes}
 \label{tab:allplanes}
 \begin{center}
 \begin{tabular}{@{}lcc}
 \hline 
Name & LGP1$^{\mathrm{mod}}$ & Great Northern Plane \\
 \hline
$r_0 \begin{pmatrix} x \\ y \\ z \end{pmatrix}$\ [kpc] & $\begin{pmatrix}   -52.1 \pm  4.6 \\ -242.1 \pm  3.2 \\ -51.4 \pm  5.5  \end{pmatrix}$  & $\begin{pmatrix}  -306.7 \pm  9.8 \\ -928.6 \pm  8.0 \\  935.0 \pm  17.5 \end{pmatrix}$ \\ 
$n \begin{pmatrix} l \\ b \end{pmatrix}$\ [$^{\circ}$] & $\begin{pmatrix}   225.4 \\ -20.8 \end{pmatrix}$  & $\begin{pmatrix}  197.6 \\ -31.5 \end{pmatrix}$ \\ 
$\Delta n$\ [$^{\circ}$] & 0.6 & 0.6 \\ 
$D_{\mathrm{MW}}$\ [kpc] & $ 182.7 \pm 2.3$ & $ 162.6 \pm 5.6$ \\ 
$D_{\mathrm{M31}}$\ [kpc] & $ 247.2 \pm 5.8$ & $ 143.8 \pm 9.2$ \\ 
$\Delta$\ [kpc] & $44.5 \pm 2.2$ & $53.4 \pm 1.5$ \\ 
$c/a$ & $0.062 \pm 0.003$ & $0.098 \pm 0.004$ \\ 
$b/a$ & $0.497 \pm 0.009$ & $0.762 \pm 0.067$ \\ 
$N_{\mathrm{members}}$ & 9 & 8 \\ 
 \hline
 \end{tabular}
 \end{center}
 \small \medskip
Parameters of the planes fitted to the LG dwarf galaxies, as discussed in Sects. \ref{sect:perseus} and \ref{sect:ngc3109leoplane}. These are:
$\mathbf{r}_0$: $x$-, $y$- and $z$-position of the centroid of the plane in the coordinate system introduced in \citet{2013MNRAS.435.1928P}.
$\mathbf{n}$: The direction of the normal vector (minor axis) of the best-fit plane in Galactic longitude $l$\ and latitude $b$.
$\Delta_{\mathrm{n}}$: Uncertainty in the normal direction. This and all other uncertainties were determined by varying the galaxy positions within their uncertainties and then determining the standard deviation in the resulting plane parameters.
$D_{\mathrm{MW}}$\ and $D_{\mathrm{M31}}$: offset of the planes from the MW and M31 position.
$\Delta$: RMS height of the galaxies from the best-fit plane.
$c/a$\ and $b/a$: ratios of the short and intermediate axis to the long axis, determined from the RMS heights in the directions of the three axes.
$N_{\mathrm{members}}$: Number of galaxies associated with the planes used for the fitting. In particular LGP1 and LGP2 might have additional satellite galaxies as members, but these were not included in the plane fits compiled here.
\end{table}

Recently, \citet{2013arXiv1310.4170M} reported the discovery of a dSph galaxy in
the vicinity of Andromeda, Persus I. At a distance of 374\,kpc from M31 it is a non-satellite galaxy according to the categorisation of \citet{2013MNRAS.435.1928P}, which considers only galaxies closer than 300 kpc, i.e. within the typically assumed virial radii of the MW and M31, to be satellites. All but one of the 15 previously known non-satellite galaxies in the Local Group were found to be close to one of two thin, highly symmetric LG planes (LGP1 and LGP2). Is Perseus I a member of one of the two non-satellite galaxy planes in the LG?

To determine whether Perseus I lies close to one of the two LG planes as reported by \citet{2013MNRAS.435.1928P} (see for example their table 3), we adopt the position and distance modulus of Perseus I from table 1 of \citet{2013arXiv1310.4170M}. In the Cartesian coordinate system of \citet{2013MNRAS.435.1928P}, this places Persues I at $(x,y,z) = (460, 94, -68)$\,kpc, with a position-uncertainty along the line connecting Perseus I and the Sun of 65\,kpc.

Perseus I indeed lies in the vicinity of one of the planes, LGP1 (offset by $141 \pm 15$\,kpc, compared to the plane's root-mean-square [RMS] height of $55$\,kpc), as defined by the dwarf galaxies UGC 4879, Leo A, Leo T, Phoenix, Tucana, WLM, Cetus, IC 1613 and Andromeda XVI. All other dwarf galaxy planes, in particular LGP2, the second non-satellite galaxy plane in the LG, are more distant than $\approx 250$\,kpc.

While Andromeda XVI is considered a member of LGP1 for formal reasons in \citet{2013MNRAS.435.1928P}, this galaxy is perfectly aligned with the GPoA (offset of only $8 \pm 3$\,kpc), its line-of-sight velocity shows that the galaxy follows the co-orbiting trend of the other GPoA members and it is at a distance of only 323\,kpc from M31. It is therefore more likely that Andromeda XVI belongs to the GPoA rather than the LGP1. Removing it from the plane-fit results in a RMS height of $\Delta = 38 \pm 2$\,kpc; short-to-long axis ratio of $c/a = 0.050 \pm 0.003$\ and intermediate-to-long axis ratio of $b/a = 0.422 \pm 0.004$; offset from MW of $D_{\mathrm{MW}} = 183.6 \pm 2.1$\,kpc and from M31 of $D_{\mathrm{MW}} = 209.9 \pm 4.4$\,kpc. The normal to the best-fit plane points to $(l,b) = (223^\circ, -22^\circ)$. Of the galaxies within 300\,kpc of M31, Triangulum/M33 and its potential satellite Andromeda XXII are both very close to the best-fit plane ($11.2 \pm 5.0$\ and $22.2 \pm 15.7$\,kpc, respectively). Perseus I is at a considerably smaller offset ($100 \pm 14$\,kpc) from this plane fit than from the one including Andromeda XVI.

This warrants inclusion of the galaxy in the modified LGP1 sample (LGP1$^{\mathrm{mod}}$), which now consists of UGC 4879, Leo A, Leo T, Phoenix, Tucana, WLM, Cetus, IC 1613 and Perseus I. The resulting parameters for LGP1$^{\mathrm{mod}}$\ are compiled in Tab. \ref{tab:allplanes}. They are similar to the fit without Perseus I: RMS height of $\Delta = 45 \pm 2$\,kpc; short-to-long axis ratio of $c/a = 0.062 \pm 0.003$\ and intermediate-to-long axis ratio of $b/a = 0.497 \pm 0.009$; offset from MW of $D_{\mathrm{MW}} = 182.7 \pm 2.3$\,kpc and from M31 of $D_{\mathrm{MW}} = 247.2 \pm 5.8$\,kpc. The normal to the best-fit plane points to $(l,b) = (225^\circ, -21^\circ)$ which is only 5 degree inclined relative to the normal of the original LGP1 ($[l,b] = [220^\circ, -22^\circ]$). M33 and Andromeda XXII both remain close to the best-fit plane ($48.6 \pm 6.1$\ and $23.6 \pm 10.6$\,kpc, respectively).

As can be seen in Fig. \ref{fig:LGplanes}, which shows an edge-on view of the LG planes LGP1$^{\mathrm{mod}}$\ and LGP2, Perseus I is clearly aligned with this thin plane. The newly discovered galaxy therefore confirms the finding by \citet{2013MNRAS.435.1928P} that essentially all non-satellite galaxies of the LG are confined to two very thin planes. Perseus I is offset by only $63 \pm 6$\,kpc from the best-fit plane, which is to be compared with the largest extent of LGP1$^{\mathrm{mod}}$\ of 2.2 Mpc between Tucana and UGC 4879. The offset might in fact be a weak indication of a bending of the plane. As seen in Fig. \ref{fig:LGplanes}, Perseus I and other nearby galaxies are offset to the right from the line indicating the edge-on view of the best-fit plane in Fig. \ref{fig:LGplanes}, while those at the top and bottom are offset to the left.

The significance of a satellite galaxy plane can be tested by comparing the observed distribution with an expected one, which in most cases is assumed to be isotropic. Due to the existence of a preferred axis in the LG (the MW--M31 line), this comparison can not be easily adopted for planes spanning the whole LG. A proper determination of the plane significance therefore requires an expected model for the distribution of the non-satellite galaxies in the LG, which is not available. Furthermore, such a test has to take observational biases like the sky coverage of surveys searching for LG dwarf galaxies into account. Due to the very inhomogeneous nature of the galaxy data, this is currently not feasible. We nevertheless try to get a rough estimate for how likely it is to find two similarly thin planes of non-satellite galaxies in the LG. Assuming that Andromeda XVI is part of the GPoA, we use the following 15 LG dwarf galaxies for this test: Andromeda XVIII, Andromeda XXVIII, Aquarius, Cetus, IC 1613, Leo A, Leo T, NGC 6822, Pegasus dIrr, Perseus I, Phoenix, Sagittarius dIrr, Tucana, UGC 4879 and WLM. Instead of assuming a model for the expected galaxy distribution, we generate 1000 randomised realisations by rotating each of the dwarf galaxy positions by individual random angles around the MW--M31 axis. This preserves their distances from both the MW and from M31. For each realisation, we then proceed as follows. Ignoring one of the LG galaxies (analogously to ignoring the Pegasus dIrr galaxy which is neither part of LGP1$^{\mathrm{mod}}$\ consisting of 9 dwarf galaxies nor of LGP2 consisting of 5 dwarf galaxies), we split the remaining galaxies into two samples, constructing all possible combinations of 9:5 objects. Planes are then fitted to both samples for each of these combinations. Both planes' RMS heights $\Delta$\ and axis ratios $c/a$\ are recorded for all 30030 possible combinations for each of the 1000 realisations. Due to the large number of possible combinations we refrain from varying the observed galaxy distances within their uncertainties and only use the most-likely values.

We then test how many of the randomised galaxy distributions contain planes which are similarly thin ($\Delta \leq 45$\,kpc) or have similar axis ratios ($c/a \leq 0.062$) as the observed LGP1$^{\mathrm{mod}}$. The mean of the minimum RMS height for the 9-galaxy combinations over all realisations is 83 kpc, with a standard deviation of 22 kpc, while the mean minimum axis ratio is 0.140, with a standard deviation of 0.037. Thus, on average a plane of galaxies as narrow as the LGP1$^{\mathrm{mod}}$\ is not expected. Of the 1000 realisations, 29 (2.9 per cent) contain a group of 9 galaxies which can be fitted by a plane with a RMS height at least as small as that of LGP1$^{\mathrm{mod}}$. Only 10 (1.0 per cent) can be fitted by a plane with a sufficiently small axis ratio. If we simultaneously require that five out of the remaining six galaxies can be fitted by a plane having a similar RMS height or axis ratio as LGP2 ($\Delta \leq 66$\,kpc, $c/a \leq 0.110$; \citealt{2013MNRAS.435.1928P}), a situation similar to the one observed in the LG occurs in only 13 (1.3 per cent, RMS height criterion) or three (0.3 per cent, axis-ratio criterion) of the 1000 random realisations. This test demonstrates that the LG planes are unexpected. We have not tested for the symmetry of the planes and their alignments, and varying the galaxy positions by only rotating them around the MW--M31 axis has a large chance of preserving information on the possible LG planes because they are parallel to the MW--M31 line. Our results should therefore be considered as upper limits.

\section{Predicted velocity dispersions for LG dSphs}
\label{sect:MOND}

In addition to the structures that dwarf galaxies trace in the LG,
their internal kinematics are also of interest. These objects are 
generally inferred to be dark matter dominated, though there was no
reason to anticipate this a priori. LCDM models can be constructed 
to match this observation, but do not provide the ability to predict
the velocity dispersion of any particular dwarf.  In contrast,
it is possible to predict a dwarf's velocity dispersion given its
photometric properties using MOND.

We employ here the method described by \citet{2013ApJ...766...22M}. Being quite remote from M31
(374\,kpc), Perseus I is well into the isolated deep MOND regime for which
the characteristic velocity dispersion follows directly from the stellar mass
($\sigma_{iso} = (4 a_0 G/81) M_*^{1/4}$). 
This makes it one of the best test cases among the dwarfs of Andromeda. 
Most (though not all) of the other dwarfs are in the regime dominated
by the external field effect so that the predicted velocity dispersions
are less certain as they depend on the properties of M31 as well as those
of the dwarfs themselves \citep[see][]{2013ApJ...766...22M}.

Given the luminosity reported by \citet{2013arXiv1310.4170M},
we predict that Perseus I should have a velocity dispersion of
$\sigma = 6.5^{+1.2}_{-1.0} \pm 1.1\;\mathrm{km}\,\mathrm{s}^{-1}$.
For consistency with \citet{2013ApJ...766...22M}
we assume a stellar mass-to-light ratio of $2\;\mathrm{M_{\odot}}/\mathrm{L_{\odot}}$,
The first uncertainty reflects a factor of two variation in mass-to-light ratio
while the second propagates the stated observational uncertainties.  
Predictions of this type have proven largely successful so far
\citep{2013ApJ...775..139M}; Perseus I provides another opportunity to test 
this a priori prediction of MOND.

There exist two other dwarf spheroidals in the Local Group that are far
removed from both the Milky Way and Andromeda: Cetus and Tucana.  Being
far from major perturbers, they should also be in the isolated MOND regime,
and provide correspondingly good tests.  However,
they tend to be overlooked since they are not grouped together with
the dwarfs that are obvious satellites.
Applying the same procedure described above given the luminosities and most-likely distances tabulated by
\citet{2012AJ....144....4M}, we predict for Cetus
$\sigma = 8.2^{+1.5}_{-1.3} \pm 0.4\;\mathrm{km}\,\mathrm{s}^{-1}$
and for Tucana $\sigma = 5.5^{+1.0}_{-0.9} \pm 0.4\;\mathrm{km}\,\mathrm{s}^{-1}$.
As before, the first uncertainty represents the range of mass-to-light ratios
$2^{+2}_{-1}\;\mathrm{M}_{\odot}/\mathrm{L}_{\odot}$.  The second uncertainty
represents the effect of the stated uncertainty in luminosity on the velocity dispersion
for the nominal assumed mass-to-light ratio of $2\;\mathrm{M}_{\odot}/\mathrm{L}_{\odot}$.
Any systematic error in distance will have a strong effect, since $L \propto D^2$.

The predicted velocity dispersions of both Cetus and Tucana compare poorly with 
observed values.  \citet{2007MNRAS.375.1364L} observe
$\sigma_{Cet} = 17\pm 2\;\mathrm{km}\,\mathrm{s}^{-1}$ for Cetus and
\citet{2009A&A...499..121F} measure 
$\sigma_{Tuc} = 15.8^{+4.1}_{-3.1}\;\mathrm{km}\,\mathrm{s}^{-1}$ for Tucana.
The observed values are a factor of $\sim 2$ and 3 higher than predicted,
respectively.  In terms of formal significance, the observed velocity dispersions
are $3.6 \sigma$ (Cetus) and $3.0 \sigma$ (Tucana) above the predicted range.

The velocity distributions of Cetus and Tucana are not particularly 
well described as Gaussians, so it is not obvious how to interpret their fitted
velocity dispersions and uncertainties.  The errors on the velocities of individual stars
are typically $\sim 8\;\mathrm{km}\,\mathrm{s}^{-1}$, so the velocity
dispersions predicted here would not be resolved: improved observations are warranted.
Nevertheless, in the absence of systematic errors, either in the overestimation
of the velocity dispersion or the underestimation of the luminosity, these
two dwarfs are problematic for MOND.

After the above text was written but shortly before we submitted this paper, 
\citet{2014arXiv1401.1208K} reported new observations of Cetus.  They measure a velocity
dispersion of $8.3 \pm 1.0\;\mathrm{km}\,\mathrm{s}^{-1}$. This agrees well with
our prediction of $8.2^{+1.5}_{-1.3} \pm 0.4\;\mathrm{km}\,\mathrm{s}^{-1}$.

\section{The NGC 3109 association and the LG}
\label{sect:NGC3109association}

\begin{figure*}
   \centering
   \includegraphics[width=140mm]{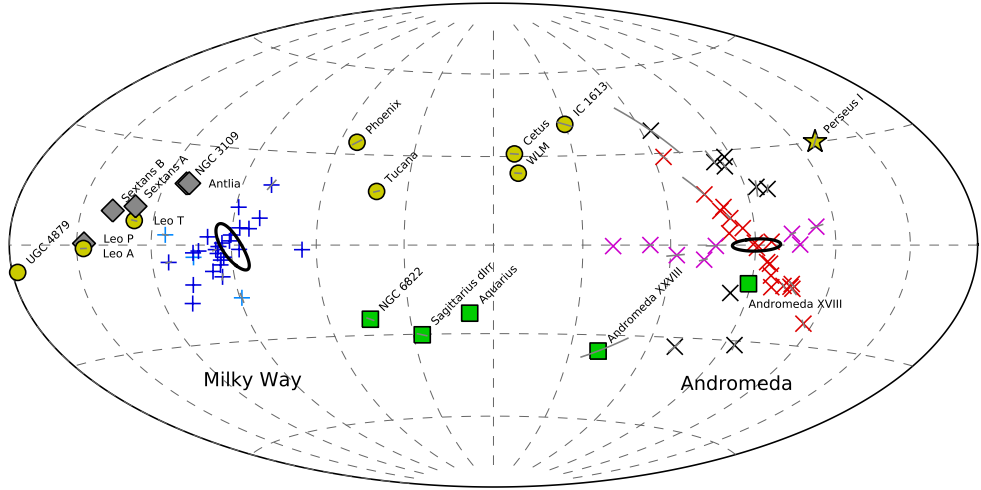}
   \caption{The positions of Local Group galaxies in an all-sky plot as seen from the midpoint between the MW and M31 (same symbols as in Fig. \ref{fig:LGplanes}). Compared to Fig. 6 in \citet{2013MNRAS.435.1928P} this plot also includes the newly discovered dwarf galaxy Perseus I and the members of the linear NGC 3109 association (grey diamonds), which align with the LGP1$^{\mathrm{mod}}$\ members Leo T, Leo A and UGC 4879, indicating that they lie in a common plane which passes through the centre of the LG. 
   }
              \label{fig:ASP}
\end{figure*}

The dwarf galaxy Leo P has recently been discovered in the vicinity of the LG by the ALFALFA survey \citep{2013AJ....146...15G,2013AJ....145..149R}. It can be considered to be a member of the NGC 3109 association of galaxies \citep[see e.g.][]{1999ApJ...517L..97V,2006AJ....132..729T}, consisting of NGC 3109, Antlia, Sextans A and Sextans B. \citet{2013arXiv1310.6365B} have shown that Leo P's position and velocity is consistent with it belonging to the highly elongated, essentially linear association, as was also noted by \citet{2013AJ....146..145M}. Given the existence of correlated satellite galaxy planes and the recent discovery that essentially all LG dwarf galaxies are confined to one of two thin and highly symmetric planes \citep{2013MNRAS.435.1928P}, it it worthwhile to investigate how the NGC 3109 association relates to these structures.

\subsection{Orientation of the NGC 3109 association}

In the following, we use the galaxy positions as compiled by \citet{2012AJ....144....4M}. However, like \citet{2013arXiv1310.6365B} we use the homogeneous set of tip of the red giant branch (TRGB) distance moduli from \citet{2009ApJS..183...67D} for the members of the NGC 3109 association. For Leo P, we adopt the recent TRGB distance modulus by \citet{2013AJ....146..145M}, according to which the galaxy is at a distance of $1.72^{+0.14}_{-0.40}$\,Mpc from the Sun.

The galaxies in the NGC 3109 association all have similar Galactocentric distances of 1.3--1.4\,Mpc, and Leo P is also consistent with this distance range. The association is therefore oriented approximately perpendicular to our the line-of sight, the angle between the long axis of the association and the line connecting the associations centroid with the position of the MW is $72^{\circ}$. The galaxies also have very similar distances from M31 and from the center of the LG, such that the association is also oriented approximately perpendicular as seen from those points (85 and $80^{\circ}$, respectively). The NGC 3109 association is therefore almost perpendicular from the line connecting its centroid with the position of M31. 

We have determined the associations's orientation in the same Cartesian coordinate system used previously by employing the method used in \citet{2013MNRAS.435.1928P}. This effectively fits an ellipsoid to the points by determining the eigenvectors of the moments of inertia tensor defined by the non-mass-weighted galaxy positions. This gives the orientations of the major, intermediate and minor axes and the root-mean-square (RMS) heights of the distribution along these axes. The resulting RMS axis ratios of the NGC 3109 association are indicative for a very elongated distribution which is extremely flat in one direction (short-to-long axis ratio $c/a = 0.014 \pm 0.007$) and a bit more extended along the intermediate axis ($b/a = 0.129 \pm 0.02$, still narrow compared to the long axis), like a ruler. The long axis points to $(l,b) = (319^\circ, -46^\circ)$\ or $(139^\circ, 46^\circ)$, with an uncertainty of $7^{\circ}$. The association therefore aligns (being only $3^\circ$\ inclined) with the Supergalactic plane \citep{1991rc3..book.....D}. It is also almost parallel to LGP1$^{\mathrm{mod}}$\ ($12^{\circ}$\ inclined), but offset by 0.3 (Leo P) to 0.5 Mpc (Antlia and NGC 3109), as can be seen in the upper panel of Fig. \ref{fig:LGplanes}. 

The LG is the closest major galaxy group to the NGC 3109 association and none of the other nearby galaxy groups listed in \citet{2009A&A...499..385P} are well aligned with the line defined by the main axis of the NGC 3109 association. The closest alignment, at an angle of $\approx 25^\circ$, is found for the M 81 group, but at a distance of $\approx 3.5$\,Mpc from the association, two times as distant as the LG, this is most likely a chance alignment.

The short axis (normal direction if it were a plane) of the ellipsoid describing the NGC 3109 association points to $(l,b) = (230^\circ, 1^\circ)$\ and $(50^\circ, -1^\circ)$.
This is extremely close to the pole of the Supergalactic Plane ($[l,b] = [47^\circ.4, 6^\circ.3]$) and also similar to the normal to the LGP1$^{\mathrm{mod}}$\ pointing to $(l,b) = (225^\circ, -21^\circ)$. Intriguingly, the normal to the GPoA has a similar orientation as well ($[l,b] = [206^\circ, 8^\circ]$, \citealt{2013MNRAS.435.1928P}), even though it is defined by the co-orbiting satellite galaxies of M31 on the opposite side of the MW than the NGC 3109 association. However, the MW is offset by 0.8\,Mpc and M31 by 0.9\,Mpc from the NGC 3109 association along this short axis direction. Thus, while some hints for a connection exist it is not immediately obvious whether the NGC 3109 association is related to the dwarf galaxy planes in the LG. In the following we will discuss additional indications why it might be part of the LG and its dwarf galaxy geometry.

\subsection{Possible origins of the association's alignment}

\citet{2013arXiv1310.6365B} mention several types of possible origins of the alignment of the NGC 3109 association. These include a tidal encounter with the MW which has stretched a pre-existing group of dwarf galaxies along its orbital plane, the formation of the galaxies as phase-space correlated TDGs, or the formation of the NGC 3109 association in a thin and cold cosmological filament which is just now starting to fall towards the LG. Here we discuss these suggestions in more detail.

\subsubsection{Common dark matter halo or infalling filament?}

Assuming the NGC 3109 association to be bound and spherically symmetric, \citet{2013arXiv1310.6365B} estimate that it would have to have a mass of $M = 3.2 \times 10^{11} M_{\sun}$\footnote{A spherical approximation is certainly a non-ideal assumption for such an extremely linear association.}. This is a significant fraction (up to one third) of the total mass of $\approx 1.0~\mathrm{to}~2.4 \times 10^{12} M_{\sun}$\ currently estimated for the MW halo \citep{2013ApJ...768..140B}. If the NGC 3109 association were embedded in such a massive dark matter halo, this halo should significantly influence the dynamics of the LG. In addition, a past interaction with the massive dark matter content of the association would have likely lead to it merging with the MW because dynamical friction must be significant in this mass range. A spherical dark matter halo encompassing the whole association would furthermore have to have a radius of at least 600 kpc, half the associations diameter. This is twice as much as the virial radii assumed for the MW and M31 and would almost reach Leo A at a distance of about 800\,kpc from the NGC 3109 association. We therefore deem it unlikely that the NGC 3109 association is a distinct gravitationally bound entity.

\citet{2013arXiv1310.6365B} suggest a cosmological accretion scenario in which the NGC 3109 association is a ``thin and cold cosmological filament'', assuming that the galaxies in the association have recently left the Hubble flow and are currently falling into the LG for the very first time. However, simulated dark matter filaments around MW-like haloes today have diameters which are larger than the virial radii of the major haloes they feed \citep{2011MNRAS.416.1377V}. Due to its much more narrow configuration, the NGC 3109 association cannot be identified with such a major dark matter filament. The dark matter hypothesis does, however, allow the freedom to propose that the NGC 3109 association is embedded in a dark matter ``sub-filament'', which would have to be oriented almost perpendicular to its closest major mass concentration. While cosmological simulations might be searched for the existence of thin sub-filaments of this kind, as of now such a suggestion unfortunately remains non-testable via observations. However, one wonders how such a dark matter filament in the vicinity of the major dark matter halo potentials of the MW and M31 could have remained essentially a straight line over a length of 1.2\,Mpc without being aligned towards the mass concentration.

The interpretation that the association is a filament falling towards the LG is in conflict with the orientation of the NGC 3109 association being almost perpendicular to the direction towards the MW or the LG barycenter. Simulated filaments of cold gas, which would be narrow enough to accommodate the thin association, point towards the central galaxy and are considered to be important at high redshifts of $z \gtrsim 2$\ only \citep[e.g.]{2009Natur.457..451D,2013arXiv1307.2102G}.
More importantly, the NGC 3109 association does not fall towards but recedes from the MW with $\approx 170\,\mathrm{km\,s}^{-1}$, a velocity similar to those of galaxies at larger distances of 1.6 to 2.2 Mpc (see e.g. fig. 5 of \citealt{2012AJ....144....4M}). The association appears to be situated beyond the zero-velocity surface of the LG ($\approx$\ 1 to 1.5\,Mpc) and therefore should follow the Hubble flow. If it follows the Hubble flow, the galaxies should never have been close to the LG (except during the Big Bang). However, the members of the NGC 3109 association appear to recede faster than the expected Hubble flow velocity at their distance (see e.g. figure 6 of \citealt{2012MNRAS.426.1808T}). If this is the case and if they are on radial orbits (otherwise their motion is even faster), the galaxies must have been close to the MW in the past.

\subsubsection{A past encounter with the LG?}
\label{sect:timing}

Assuming that all galaxies in the NGC 3109 association are receding on radial orbits from the MW (i.e. the tangential velocity is zero), and that they have been travelling with their current Galactocentric velocity (no acceleration) gives a rough estimate for their travel time. If we transform the measured Heliocentric velocities to Galactocentric ones by assuming a circular velocity of the local standard of rest (LSR) of 220 km/s and a peculiar motion of the Sun as measured by \citealt{2010MNRAS.403.1829S}, the travel times are $6.7 \pm 0.1$\,Gyr for NGC 3109, $8.2 \pm 0.2$ Gyr for Sextans B, $8.6 \pm 0.2$\,Gyr for Antlia and $8.6 \pm 0.3$\,Gyr for Sextans A. Due to the large distance uncertainty for Leo P its travel time is in the range of 7.6 to 10.6 Gyr.  The stated uncertainties are based on the distance uncertainties and an assumed uncertainty in the line-of-sight velocities of the galaxies of 2 km\,s$^{-1}$. Of course, the major inaccuracy comes from the over-simplified assumption of radial orbits without acceleration. Another major source of uncertainty is the LSR velocity. If we assume a LSR velocity of 240 km\,s$^{-1}$, instead of the previously used 220 km\,s$^{-1}$, the travel times become 0.75 Gyr longer on average.

If the galaxies in the NGC 3109 association do not have a significant tangential velocity, this timing estimate demonstrates that they have been close to the MW in the past. More importantly, the galaxies are consistent with all having been nearby at about the same time, $\approx 8.5$\,Gyr ago. Only NGC 3109's timing appears to be off by 1.5--2\, Gyr, but unaccounted-for tangential motion, gravitational acceleration by the LG galaxies and others and the possible interaction between members of the NGC 3109 association make this estimate unreliable on such time scales.

That the NGC 3109 members have been close to the MW at a similar time in the past is also consistent with results by \citet{2013MNRAS.tmp.2500S}. In their model of the dynamical history of the LG, the NGC 3109 association members had a close passage (pericenter $\approx 25$\,kpc) with the MW about 7\,Gyr ago. However, at such a distance dynamical friction, in particular for massive objects such as NGC 3109 or the putative dark matter halo needed if one requires that the association as a whole is gravitationally bound, cannot be neglected. As \citet{2013MNRAS.tmp.2500S} do not consider dynamical friction in their study, the orbit of the association's members as derived from their study becomes unreliable as soon as they come close to the MW halo.

Another independent indication that the members of the NGC 3109 association have been close to the LG before is their identification as probable backsplash galaxies, galaxies which have passed through the virial volume of the MW in the past but are now situated outside of it. By comparing the distances and line-of-sight velocities of LG galaxies with sub-haloes in the Via Lactea II simulation, \citet{2012MNRAS.426.1808T} identify all four of the then-known galaxies in the NGC 3109 association as likely backsplash haloes that have interacted with the MW before. 

It therefore appears likely that the NGC 3109 association had a past encounter with the MW, in conflict with the interpretation as an infalling dark matter filament. A scenario in which a pre-existing group of dwarf galaxies collided with the MW and was tidally stretched along its orbit appears plausible, but the effects of dynamical friction in such a scenario will need to be assessed. It is currently not possible to discriminate such a scenario from one in which the galaxies were formed as TDGs in a past encounter in the LG, but there are additional coincidences which make a TDG origin look more appealing.

\subsubsection{Consistent with a TDG origin?}

As shown before, given their current distances and line-of-sight velocities, the galaxies in the NGC 3109 association are consistent with having been in the vicinity of the MW at roughly the same time about 7 to 9 Gyr ago. This leads us to the other possible origin of the association discussed by \citet{2013arXiv1310.6365B}: the galaxies might have been expelled as TDGs in a galaxy encounter. Interestingly, the estimated travel times are in good agreement with independent estimates for the timing of major galaxy encounters suggested as the origin of a population of TDGs in the LG:
   \begin{itemize}
      \item Based on the ages of the young halo globular clusters, the VPOS around the MW was estimated to have formed 9--12\,Gyr ago \citep{2012MNRAS.423.1109P}.
      \item A major merger might have formed M31 8--9\,Gyr ago. This event must have produced TDGs, some of which would constitute the GPoA today while others would have been expelled towards the MW, maybe forming the VPOS \citep{2010ApJ...725..542H,2012MNRAS.427.1769F,2013MNRAS.431.3543H}. 
      \item In MOND the observed baryonic masses of the MW and M31 and their relative velocity require a past encounter between the two major galaxies to have happened 7--11\,Gyr ago \citep{2013A&A...557L...3Z}.
   \end{itemize}

The NGC 3109 association is therefore consistent with having formed in the same major galaxy encounter and thus such an origin cannot be ruled out at present. The SFHs of the non-satellite galaxies do not provide conclusive information either, because most non-satellite galaxies in the LG are dIrrs which show ongoing star formation, consistent with expectations for gas-rich TDGs if the present high-resolution simulations of TDG formation and early evolution can be extrapolated to many Gyr \citep{2007A&A...470L...5R, 2013arXiv1311.2932P}. Differing environmental effects will furthermore diversify the evolutionary histories of TDGs born in the same event over time. The apparent dark matter content of some of these galaxies, unexpected for TDGs in a cold dark matter universe \citep[but nevertheless observed in young TDGs, see][]{2007Sci...316.1166B,2007A&A...472L..25G}, is not a conclusive argument against the possible TDG nature of the NGC 3109 association either, as inflated velocity dispersions are expected in MOND as exemplified in Sect. \ref{sect:MOND}.

To be consistent with a scenario in which a major fraction of the LG galaxies was formed in one common major encounter, the members of the NGC 3109 association must show signs of being related to the other dwarf galaxy structures known in the LG, in particular the VPOS, the GPoA and potentially the LG dwarf galaxy planes, because TDGs form as a phase-space correlated population, i.e. in a highly flattened tidal tail. If the NGC 3109 association shows alignments with other LG dwarf galaxies, this might indicate that the NGC 3109 association is, contrary to current belief, not an isolated entity and might share a common origin with other dwarf galaxy structures.

\subsection{The NGC 3109 association as an extension of LGP1}
\label{sect:ngc3109leoplane}

Given that the total extent of the NGC 3109 association (the distance between Leo P and Antlia) is 1.2 Mpc, while the three LGP1 members UGC 4879, Leo A and Leo T are at distances of only 0.8--1 Mpc from their respectively closest member of the NGC 3109 association, these eight galaxies might well constitute one common structure of square-megaparsec size. The difference in the Galactocentric line-of-sight velocities between the NGC 3109 association (receding) and the other three galaxies (slowly approaching) might then be simply due to the stronger gravitational attraction acting on the more nearby galaxies. Indeed, as seen from the center of the LG the members of the NGC 3109 association align along a similar band as defined by UGC 4879, Leo A and Leo T (see Fig. \ref{fig:ASP}). This indicates that they are in a common plane running through the center of the LG. Fitting a plane to the eight galaxy positions confirms that the galaxies are confined to a thin planar structure, which we will refer to as the \textit{Great Northern Plane}. The parameters of the best-fit plane are compiled in Tab. \ref{tab:allplanes}. The plane has a normal vector pointing to $(l, b) = (197.6^\circ, -31.5^\circ)$\ and RMS axis ratios of $c/a = 0.098 \pm 0.004$\ and $b/a = 0.762 \pm 0.067$. As expected, it runs through the center of the LG (see lower panel in Fig. \ref{fig:LGplanes}). The galaxy with the largest offset of only 86\,kpc from the best-fit plane is UGC 4879. Compared to LGP1$^{\mathrm{mod}}$, with which it shares three members, the plane has a similar offset from the MW ($162.6 \pm 5.6$\,kpc) and a similar RMS height ($\Delta = 53.4 \pm 1.5$\,kpc). There is no other known dwarf galaxy in the northern hemisphere of the MW and between Galactocentric distances of 0.3 to 1.7 \,Mpc from the MW, such that all currently known nearby non-satellite dwarf galaxies to the north of the MW are confined to this one common plane.

To determine how significant this planar alignment is, we follow the same principle as in Sect. \ref{sect:perseus} (thus the same caveats apply here). We randomise the positions of the eight non-satellite galaxies in the northern hemisphere of the MW and then determine the RMS height of the planes fitted to these new positions. We generate $10^5$\ realisations for each of two randomisation methods. The first method draws the galaxy positions randomly from an isotropic distribution, but we force the galaxies to be  confined to $b \geq 20^{\circ}$. This make sure they are in the Northern Galactic hemisphere and in an area not obscured by the MW disc. This method conserves the distances from the MW, but not from M31. In this case, only five (0.005 per cent) of all realizations have $\Delta \leq 53.4$\,kpc.
In the second method we instead conserve the distances from both the MW and from M31, by rotating the galaxy positions by a random angle around the MW--M31 axis. Again they are required to be at $b \geq 20^{\circ}$. In this case only seven (0.007 per cent) of the realisations result in sufficiently thin best-fit planes. However, Antlia might be interpreted as a satellite of NGC 3109 \citep{1999ApJ...517L..97V}, such that these two galaxies might not be treated as two independent systems. If we exclude Antlia from this analysis (thus using only seven instead of eight galaxies), the probabilities to find a plane at least as narrow as the observed one increase to 0.079 per cent if the galaxy positions are chosen from an isotropic distribution and to 0.037 per cent if the galaxy positions are individually rotated by a random angle around the MW--M31 axis. A plane as narrow as the observed Great Northern Plane is therefore very unlikely to occur by pure chance.

The plane is inclined relative to LGP1$^{\mathrm{mod}}$ by only 27$^\circ$. As can be seen in the lower panel of Fig. \ref{fig:LGplanes}, which shows these two planes edge-on, the line at which they intersect passes close to the MW. Interestingly, the plane is also inclined by only 25$^\circ$ from the average orbital pole of the MW satellites with known proper motions, around which the individual satellite orbital poles scatter with a spherical standard distance of $\Delta_{\mathrm{sph}} = 29^\circ$ \citep{2013MNRAS.435.2116P}. The plane is therefore consistent with being aligned with the preferred orbital plane of the MW satellites in the VPOS, but is somewhat less polar than the VPOS. Relative to the GPoA of M31 satellites it is inclined by $40^\circ$. Interestingly, even though planes of satellite galaxies can be dynamically stable, depending on the galactic potential in which they are embedded their orientation can change due to precession and inclined planes can evolve into a more polar orientation (Klarmann et al. in prep.).

It is possible that LGP1$^{\mathrm{mod}}$ bends close to the position of the MW, such that the NGC 3109 association is in fact a part of the same structure of non-satellite galaxies in the LG. For such a causal connection to be feasible, the NGC 3109 members must have been closer to the LG in the past, which appears to be supported by the timing estimate presented before. The NGC 3109 association therefore appears to be connected to the previously-identified dwarf galaxy structures of the LG such that the LG geometry does not rule out the possibility of a TDG origin for its members.

\section{Does the search for backsplash galaxies identify TDGs?}
\label{sect:backsplash}

\begin{figure}
   \centering
   \includegraphics[width=85mm]{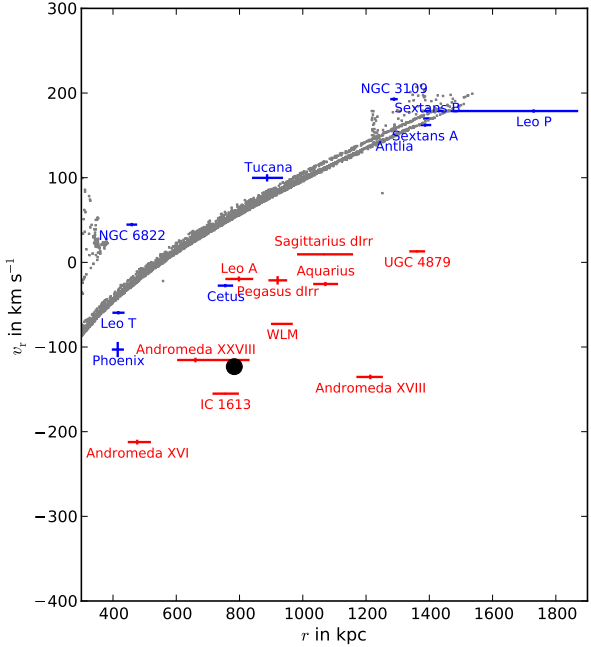}
   \caption{
   Galactocentric radial velocity $v_{\mathrm{r}}$\ of LG dwarf galaxies is plotted against their Galactocentric distance $r$. Compare to fig. 6 of \citet{2012MNRAS.426.1808T}. The black circle denotes M31. Dwarf galaxies identified as likely backsplash objects by \citet{2012MNRAS.426.1808T} are marked in blue, the others in red. Leo A has a non-negligible but below 50 per cent likelihood of being a backsplash galaxy. Leo P was discovered only after the study of \citet{2012MNRAS.426.1808T} and should also be interpreted as a backsplash candidate.
   The grey dots are tidal debris particles from a $N$-body model of a galaxy collision at about 9\,Gyr after the first pericenter. Two disc galaxies embedded in $10^{12}\,\mathrm{M}_{\sun}$\ dark matter haloes merge, with one falling in on a polar, prograde orbit. The model parameters have not been fine-tuned to reproduce the MW or the surrounding LG galaxies, but the distribution of particles illustrates that tidal debris have similar properties as the backsplash galaxy candidates in this plot. This demonstrates that TDGs might be wrongly identified as backsplash galaxy candidates. Note that more than one tidal tail can be formed if there are several pericenter passages before the final merger and that a different origin of the TDGs, e.g. in M31 or at the barycenter of the LG in case of a past MW--M31 encounter, would result in a more complex distribution of tidal debris.
   }
              \label{fig:backsplashdebris}
\end{figure}

Interestingly, of the seven galaxies in the sample by \citet{2012MNRAS.426.1808T} which lie to the north of the MW ($b > 0^\circ$), five are identified as likely backsplash galaxies and one of the remaining two (Leo A) has a non-negligible likelihood if sub-haloes close to a M31 analogue are excluded from the analysis. In the whole northern hemisphere of the MW (the southern hemisphere is more complicated due to the presence of M31 and its satellite galaxy population) the majority of dwarf galaxies are likely backsplash objects, which is unexpected from cosmological simulations. According to \citet{2012MNRAS.426.1808T}, overall only 13 per cent of all sub-haloes between 300 and 1500\,kpc from the MW should be backsplash haloes.

This can be expressed in another way, which might highlight a more curious coincidence: all galaxies in the Great Northern Plane except for UGC 4879\footnote{Including Leo P, which was not yet discovered when \citealt{2012MNRAS.426.1808T} performed their analysis but has a similar distance and line-of-sight velocity as the members of the NGC 3109 association identified as likely backsplash galaxies} are possible backsplash objects, posing the question why they should all be in the same thin, planar structure.
Even more puzzling is that most of the remaining five likely backsplash galaxies of \citet{2012MNRAS.426.1808T} situated in the southern hemisphere of the MW are also close to the Great Northern Plane defined by the northern non-satellite galaxies only: NGC 185 at a distance of 45\,kpc from the best-fit plane\footnote{NGC 185 is a satellite galaxy of M31, situated in the GPoA and its LOS velocity indicates that it is also co-orbiting in the same sense as the other members of the satellite plane. This makes it unlikely that it is a dark matter sub-halo which has been close to the MW before, as it would have to end up in the unrelated satellite structure with the right velocity by chance. This case demonstrates that it is problematic to attempt an identification of backsplash galaxies based only on their Galactocentric distance and line-of-sight velocity if one does include the galaxies in the direction of M31.}, Phoenix  at 87\,kpc\footnote{Note that the identification as a backsplash candidate by \citet{2012MNRAS.426.1808T} is based on the Heliocentric velocity for Phoenix reported in \citet{1997AJ....114.1313C} of 56\,km\,s$^{-1}$, whereas more recent measurements yield $-13 \pm 9$\,km\,s$^{-1}$\ \citep{2002MNRAS.336..643I} or even $-52 \pm 6$\,km\,s$^{-1}$ \citep{2001AJ....121.2572G}. These result in Phoenix approaching the MW much faster than assumed in the analysis by \citet{2012MNRAS.426.1808T}, such that identification as a backsplash candidate will be less likely for this galaxy.}, Tucana at 113\,kpc and maybe Cetus which is offset by 204\,kpc. Only NGC 6822 has a much larger offset of 460\,kpc. The distance uncertainties in the galaxy positions result in uncertainties in the offsets from the best-fit plane of 8 to 11\,kpc.

That all distant dwarf galaxies in the northern hemisphere of the MW are confined to one single plane, which in addition is inclined by less than $30^{\circ}$\ from both the average orbital plane of the MW satellites co-orbiting in the VPOS and from the LGP1 might be more than a lucky coincidence. The study by \citet{2012MNRAS.426.1808T}, according to which the distance and the line-of-sight velocity makes five out of seven of these galaxies more consistent with being backsplash galaxies than infalling ones, whereas the latter type should be seven times more abundant, might indicate another small-scale problem for the current $\Lambda$CDM cosmology. To find at least 5 out of 7 galaxies to be backsplash objects if only 13 per cent of all galaxies in this distance range should be backsplash objects has a likelihood of only 0.12 per cent. Even if half of the candidates are false positives, i.e. we would expect to find that 26 per cent of all galaxies are identified as candidates, the chance to draw 5 out of 7 candidates is still only 3 per cent. If Leo P is included as a backsplash object, these likeliehoods drop even more. Unless cosmological simulations can show that there is a large excess of backsplash haloes in one hemisphere from a central galaxy, the $\Lambda$CDM model will face an ``overabundant backsplash problem'' in the LG. Already now we have to wonder how, in a whole hemisphere, there can be more galaxies which have been close to the MW before and escaped to large distances than there are galaxies which are falling in, unless at least some of them have formed in and been expelled from the LG as TDGs.

In addition, if the identification as backsplash galaxies by \citet{2012MNRAS.426.1808T} is confirmed by other studies, we might even face a ``planar backsplash problem'' as almost all of the current backsplash candidates lie within about 100 kpc of the same plane, which is defined by only the northern non-satellite galaxies and measures more than 1\,Mpc in diameter. Again, this might be well understood in a TDG scenario where new galaxies form in a tidal tail as a phase-space correlated population, but there is no reason to anticipate a planar configuration among backsplash galaxies.

As illustrated in Fig. \ref{fig:backsplashdebris}, TDGs expelled from a major galaxy encounter will have similar orbital properties as backsplash-sub-haloes. They are receding from the main galaxy with high velocities on almost radial orbits because they originate from material in the galactic disc. Their formation in a common event furthermore implies that the galaxies have similar travel times, in contrast to backsplash galaxies which might have been individually accreted. This could be a promising route to observationally discriminate between the two origins.
Such a discrimination will be important because the search for backsplash galaxies in the LG might in fact be good at identifying ancient TDGs expelled in a past major galaxy encounter.

\section{Schematic structure of the Local Group}
\label{sect:scheme}

\begin{figure*}
   \centering
   \includegraphics[width=150mm]{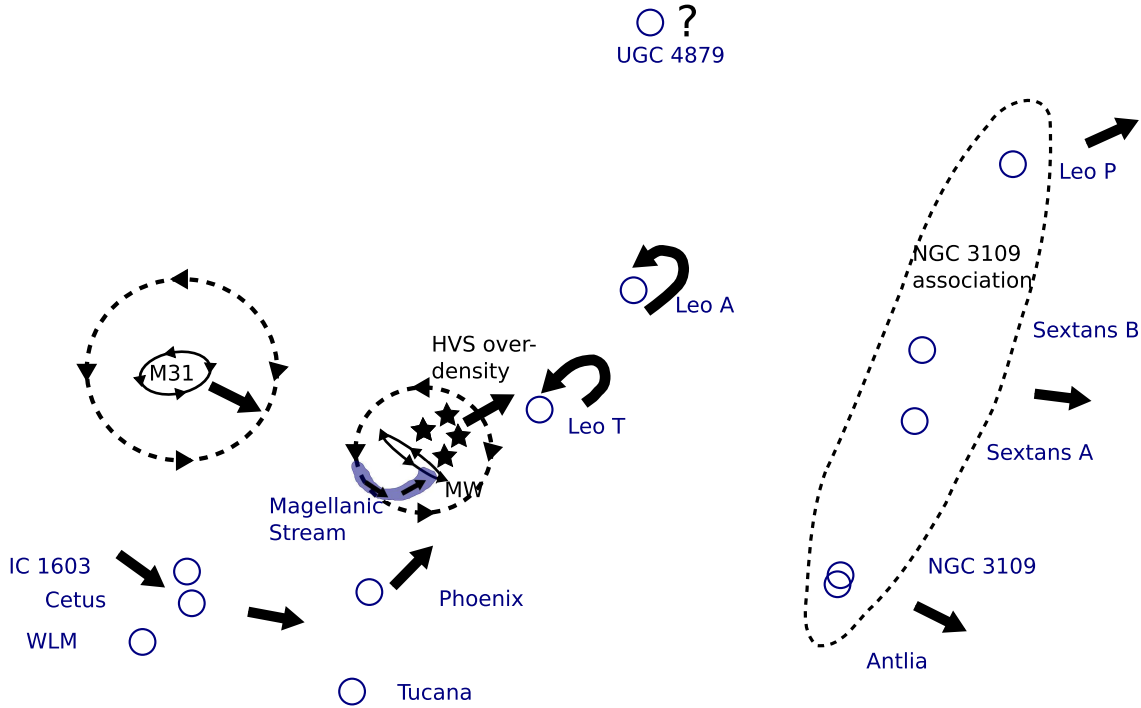}
   \caption{Schematic sketch of the LG looking approximately face-on onto the most pronounced dwarf galaxy structures. Not shown are the five non-satellite dwarf galaxies associated with the second LG plane LGP2. See Sect. \ref{sect:scheme} for a detailed description and discussion. Note that the size of the ellipses indicating the orientation of the galactic discs of the MW and M31, the circles indicating the dwarf galaxy positions, the Magellanic Stream and the HVSs are not to scale.
   }
              \label{fig:LGsketch}
\end{figure*}

The approximate alignment of the MW satellites co-orbiting in the VPOS \citep{2012MNRAS.423.1109P,2013MNRAS.435.2116P}, the M31 satellites co-orbiting in the GPoA \citep{2013Natur.493...62I}, the dominant non-satellite galaxy plane LGP1$^{\mathrm{mod}}$\ in the LG and its connection to the NGC 3109 association (this work) motivate our attempt to schematically sketch the geometry of the LG in Fig. \ref{fig:LGsketch}. It shows the before-mentioned structures approximately face-on. Starting on the left, we see the orientation of M31 (solid ellipse) and the GPoA consisting of more than half its satellite galaxies (dashed circle). The arrow heads along the circles indicate the sense of rotation.
M31 is moving towards the MW on an almost radial orbit \citep{2012ApJ...753....7S,2012ApJ...753....8V}. Towards the bottom of the sketch we see a number of dwarf galaxies belonging to LGP1$^{\mathrm{mod}}$\ which appear to form a kind of bridge between the galaxies (small circles). As seen from the MW these are also approaching, except for Tucana (which, as one of the only two distant dSphs in the LG, is also morphologically different from most of the shown dwarf galaxies which are predominantly gas-rich dIrrs). These five 'bridge'-galaxies have similar positions on the sky and similar line-of-sight velocities like the Magellanic Stream \citep{2013MNRAS.435.1928P}, which falls in towards the MW with the LMC and SMC in the southern Galactic hemisphere.

The MW itself is oriented almost edge-on in this view (flatter black ellipse), but its polar structure of satellite galaxies is seen face-on (dashed circle). We can see that both satellite galaxy structures orbit in the same sense. In the north of the MW an over-density of hypervelocity stars (HVS) recede in the direction of the constellation of Leo \citep{2012ApJ...754L...2B} and opposite to the Magellanic Stream \citep{2013MNRAS.435.1928P}, which is relevant in this context because the tidal disruption of a dwarf galaxy close to the Galactic center has been suggested as a possible formation mechanism for grouped HVSs \citep{2009ApJ...691L..63A}. Whether this over-density continues to lower Galactic latitudes is currently unknown. The two galaxies at intermediate distance in the same constellation, Leo T and Leo A both have small approaching Galactocentric line-of-sight velocities. They might just have turned around, falling back towards the MW after having passed and receded from it some time ago. This is supported by their low velocities. \citet{2013ApJ...768..140B} predict that to be on its first infall towards the MW, Leo A would have to have a large tangential velocity of two times its radial velocity. Furthermore, Phoenix on the other side of the MW is at essentially the same distance as Leo T (415 compared to 422\,kpc), but it is falling towards the MW with a twice as large Galactocentric line-of-sight velocity (-103 compared to -58\,km\,s$^{-1}$). Finally, the members of the NGC 3109 association, starting with Leo P in the same constellation mentioned before and continuing to lower Galactic latitudes, are all receding with high velocities which indicate that they have been close to the MW in the past.

Not shown in the sketch is Leo I, the most distant of the classical MW satellite galaxies. It appears not to co-orbit in the VPOS and -- in contrast to all other classical satellites -- has a larger radial than tangential velocity \citep{2013MNRAS.435.2116P}. It moves away from the MW and might even be unbound to the MW, which would be unexpected if Leo I traces a dark matter sub-halo \citep{2013ApJ...768..139S,2013ApJ...768..140B}.  Leo I's 3D velocity vector \citep{2013MNRAS.435.2116P} points into the direction of Leo T: as seen from Leo I, Leo T's position is at $(l,b) = (202^\circ, 34^\circ)$\ and Leo I's most-likely velocity vector points to $(l,b) = (192^\circ, 29^\circ)$. In the future, the galaxy might therefore become a part of the Great Northern Plane.

UGC 4879, the topmost galaxy in the sketch, does not fit in. It has a very low line-of-sight velocity and its free-fall time from its present position onto the LG is larger than a Hubble time \citep{2012AJ....144....4M}. It is therefore thought to have never interacted with a major LG galaxy \citep{2008MNRAS.387L..45K}. In addition, the galaxy has the largest offset from the Great Northern Plane (see Sect. \ref{sect:ngc3109leoplane}) and according to \citet{2011AJ....141..106J} almost all its stars appear to have formed more than 10\,Gyr ago. Finally, of all galaxies in the Great Northern Plane, UGC 4879 has the lowest gas to star mass ratio \citep{2012AJ....144....4M}. UGC 4879 therefore seems to be the most unlikely galaxy to be associated to the dwarf galaxy structure in the LG.

\begin{figure}
   \centering
   \includegraphics[width=85mm]{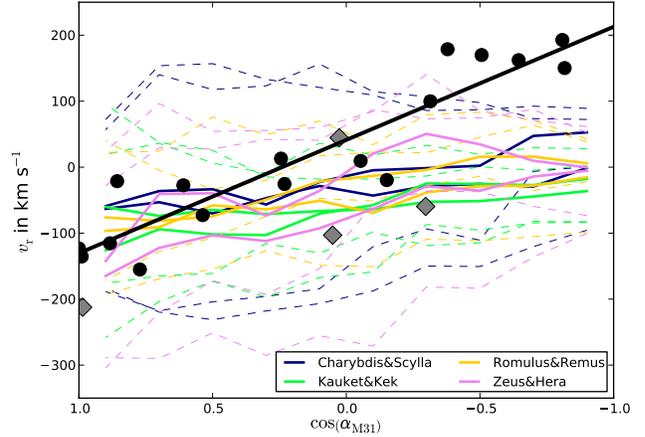}
   \caption{
   Cosine of the angle between a galaxy's position and the position of M31 on the sky, $\cos \left( \alpha_{\mathrm{M31}} \right)$, against the Galactocentric radial velocity, $v_{\mathrm{r}}$, of galaxies in the LG (within 1.8\,Mpc from the MW). Black dots represent galaxies more distant than 600\,kpc from the MW while grey diamonds represent galaxies at a distance between 300 and 600\,kpc, which might be more affected by the gravitational pull of the Galactic potential. Galaxies closer than 300\,kpc to the MW or M31, except for M31 itself, are not shown. 
   The coloured lines represent sub-halo velocities measured from one major dark matter halo in the ELVIS suite of cosmological simulations of LG equivalents \citep{2013arXiv1310.6746G}, where $\alpha$\ is measured from the position of the other major halo in the simulated group.
   The solid line is a fit to the observed galaxies at more than 600\,kpc distance from the MW resulting in $v_{\mathrm{r}} (\alpha_{\mathrm{M31}}) \approx \left( 40 - 170 \times \cos\alpha_{\mathrm{M31}} \right)\,\mathrm{km\,s}^{-1}$.
   }
              \label{fig:M31anglevel}
\end{figure}

Overall, there appears to be a trend that a variety of objects preferentially fall in from the Galactic South while those in the Galactic North recede. This is confirmed by Fig. \ref{fig:M31anglevel}, which plots the cosine of the angle between the position on the sky of a non-satellite LG galaxy and the position of M31, $\cos \left( \alpha_{\mathrm{M31}} \right)$, against the galaxy's Galactocentric radial velocity $v_{\mathrm{r}}$. For comparison, the coloured lines illustrate the average radial velocities of sub-haloes in the ELVIS suite of cosmological simulations modelling LG-like galaxy pairs \citep{2013arXiv1310.6746G}. They are measured in 10 bins in $\cos \left( \alpha \right)$, where $\alpha$\ is the angle between a sub-halo position, as seen from one major dark matter halo (representing the MW), and the position of the other major dark matter halo (representing M31). All sub-haloes within 0.3 and 1.8\,Mpc from the MW equivalent and outside of 300\,kpc from the M31 equivalent are included. The dashed lines indicate the scatter in sub-halo velocities, 95 per cent of all sub-halo velocities are below the upper dashed lines and similarly 95 per cent are above the lower. Two sets of lines are shown for each pair of galaxies, each assuming a different main halo of each halo pair to be the MW equivalent. For clarity only the four simulations with the largest volume uncontaminated by low-resolution particles are included, but the remaining 8 simulations in the ELVIS suite cover the same range in $v_{\mathrm{r}}$.
   
   Fig. \ref{fig:M31anglevel} shows that the LG galaxies tend to approach from the direction of M31 and recede in the opposite direction. If the LG were simply collapsing towards the LG barycenter, the galaxies in the direction of M31 are expected to approach but those on the opposite side are expected fall towards the LG barycenter together with the MW, resulting in radial velocities closer to zero. This is confirmed by the cosmological simulations, which at $\cos \left( \alpha_{\mathrm{M31}} \right) < -0.5$\ scatter by $\approx \pm 100\,\mathrm{km\,s}^{-1}$\ around $v_{\mathrm{r}} \approx 0\,\mathrm{km\,s}^{-1}$. The observed velocities, however, continue to larger receding values of 150 to 190\,km\,s$^{-1}$\ for the direction opposite to M31. Such high radial velocities are not found for dark matter sub-haloes in the simulations. 
While part of the velocity gradient will thus be due to the MW falling towards M31, the continuation of this trend in the direction opposite to M31 indicates that we might witness a larger-scale flow of dwarf galaxies within a flattened structure. The linear trend of increasing velocities with decreasing $\cos \left( \alpha_{\mathrm{M31}} \right)$\ can be interpreted as due to the MW moving through the LG dwarf galaxy population, such that its gravitational potential is deforming the structure (in addition to external tides, \citealt{1989MNRAS.240..195R,2009A&A...499..385P}) and the MW might have accreted or is still accreting dwarf galaxies from it (such as the LMC if it is on its first infall), while others are leaving on the opposite side after having passed close to the MW (such as the potentially unbound Leo I). While this schematic representation currently is speculative (as we do not know the tangential motions) and does not prove any scenario right, the geometry presented in Fig. \ref{fig:LGsketch} might help to constrain attempts to model the LG.

\section{Open questions}
\label{sect:openquestions}

Despite the intriguing alignments discovered among the currently known LG galaxies, it is to be expected that additional nearby galaxies will be discovered in the future. As our current knowledge is based on a large number of different sources, it is impossible to accurately consider selection effects and the possibility, however unlikely, that additional dwarf galaxies will preferentially be found outside of the structures does exist. We essentially do not know what we have not seen yet. Nevertheless, for searches for nearby dwarf galaxies, the northern Galactic hemisphere is generally considered the best-studied direction because the SDSS sky coverage initially focussed on this region. This might give some confidence in the significance of the Great Northern Plane. This is further supported by the present non-detection of new bright dwarf galaxy candidates in single-epoch data of the Pan-STARRS1 survey. While based on an approximately 1 magnitude shallower photometric depth than the SDSS, it covers a significantly larger fraction of the sky (away from the VPOS) and therefore supports the satellite galaxy anisotropy revealed by the presently known dwarf galaxies \citep{2013sf2a.conf..363L}. However, despite being covered by the SDSS, Leo P was only found by the less extended ALFALFA HI survey, indicating that additional faint galaxies might hide in the LG.

Only once proper motions of the LG dwarf galaxies are known will we be able to directly determine whether the planar structures are dynamically stable. This is so far only possible for the 11 classical MW satellites in the VPOS, of which nine are found to be consistent with moving within the structure \citep{2013MNRAS.435.2116P}, while one (Leo I) is moving towards the Great Northern Plane (as discussed in Sect. \ref{sect:scheme}). Promising developments in this regard, such as the measurement of the proper motions of the M31 satellites M33 and IC 10 via water masers \citep{2005Sci...307.1440B, 2007A&A...462..101B} or the optical proper motion measurement of the MW's most-distant classical satellite Leo I \citep{2013ApJ...768..139S} and of M31 \citep{2012ApJ...753....7S}, indicate that acquisition of three-dimensional velocity data for a larger sample of LG galaxies might be feasible in the near future.

In addition to the tentativeness of the dynamical stability of the dwarf galaxy structures, all of the different scenarios of their formation have yet to address serious issues. Most importantly, if the observed dwarf galaxies are identified with dark matter sub-haloes such as those produced in cosmological simulations, these do not naturally result in the observed planar and co-orbiting structures. Most claims of the contrary \citep{2008MNRAS.385.1365L,2008ApJ...686L..61D,2009MNRAS.399..550L,2011MNRAS.415.2607D,2013MNRAS.429.1502W} have subsequently been addressed in the literature and shown to lack important aspects of the observed situation or to be inconsistent with additional observations \citep[][Ibata et al. in prep, Pawlowski et al. in prep.]{2009ApJ...697..269M,2012MNRAS.424...80P,2013MNRAS.435.2116P}. Furthermore, the inability of present simulations to reproduce the observed structures comes in addition to the other known small-scale problems of the $\Lambda$CDM cosmology \citep[e.g.][]{2010A&A...523A..32K,2012PASA...29..395K,2013JPhCS.437a2001F,2014arXiv1401.1146W}.

In the alternative TDG scenario phase-space correlated dwarf galaxies occur naturally, but there are two crucial questions that yet lack decisive answers: Why do the observed LG dwarf galaxies have large velocity dispersions that are classically interpreted as a strong indication for dark matter? And how could TDGs, born out of pre-processed material stripped from much larger galaxies, end up on the mass-metallicity relation?
Modified gravity models \citep[e.g.][]{2007ApJ...667L..45M,2007A&A...472L..25G,2013ApJ...766...22M} and non-equilibrium dynamics \citep[e.g.][Yang et al. in prep.]{1997NewA....2..139K,2007MNRAS.376..387M,2012MNRAS.424.1941C,2013MNRAS.436..839S} might provide answers to the first, and the early formation of TDGs at a redshift of $z \approx 2$\ and out of less-metal rich material from the rim of interacting galaxies might indicate a possible approach to the second. However, none of these issues have yet been satisfactorily investigated in a full and self-consistent model forming a LG-like group of galaxies.

\section{Summary and Conclusions}
\label{sect:conclusion}

We have investigated how the recently discovered dSph galaxy Perseus I and the NGC 3109 association extended with the recently discovered dwarf galaxy Leo P are related to, and might fit in with, the dwarf galaxy structures present in the LG. 
Our work has shown that the NGC 3109 association cannot necessarily be interpreted as an independent group of galaxies, but might be related to the LG dwarf galaxy population and as such might provide important constraints on attempts to model the whole LG. 
The main results of our analysis are:

   \begin{enumerate}
      \item Perseus I is consistent with being part of the LGP1$^{\mathrm{mod}}$, the dominant plane of non-satellite galaxies in the LG, at least if Andromeda XVI is associated with the GPoA.
      \item In the context of MOND, we have predicted Perseus I's velocity dispersion to be $\sigma = 6.5^{+1.2}_{-1.0} \pm 1.1\,\mathrm{km\,s}^{-1}$. The corresponding prediction for Cetus ($\sigma = 8.2^{+1.5}_{-1.3} \pm 0.4\,\mathrm{km\,s}^{-1}$) is in much better agreement with the more recent observational value \citep[$\sigma = 8.3 \pm 1.0\,\mathrm{km\,s}^{-1}$:][]{2014arXiv1401.1208K} than with the previous measurement \citep[$\sigma = 17 \pm 2.0\,\mathrm{km\,s}^{-1}$:][]{2007MNRAS.375.1364L}. The prediction for Tucana ($\sigma = 5.5^{+1.0}_{-0.9} \pm 0.4\,\mathrm{km\,s}^{-1}$) is in conflict with the available measurement \citep[$\sigma = 15.8^{+4.1}_{-3.1}\,\mathrm{km\,s}^{-1}$:][]{2009A&A...499..121F}. We note that the observations of \citet{2009A&A...499..121F} lack the spectral resolution to resolve the predicted velocity dispersion. Unfortunately, no similar predictions are possible in a $\Lambda$CDM context.
      \item The orientation of the NGC 3109 association consisting of the dwarf galaxies NGC 3109, Antlia, Sextans A, Sextans B and Leo P has been determined in the same coordinate system used to review the planes of co-orbiting satellite galaxies around the MW and M31 and the symmetric larger-scale dwarf galaxy structure in the LG \citep{2013MNRAS.435.1928P}. The association aligns with the Supergalactic Plane, is almost perpendicular to our line-of-sight and parallel but offset by 300-500\,kpc to LGP1$^{\mathrm{mod}}$.
      \item The members of the NGC 3109 association have large receding velocities which indicate that they have been close to the MW in the past, possibly at the same time about 7--9\,Gyr ago. This is consistent with their orbits passing within $\approx 25$\,kpc of the MW suggested by \citet{2013MNRAS.tmp.2500S} and the identification as likely backsplash galaxies by \citet{2012MNRAS.426.1808T}. Together with the association's extremely narrow extent and perpendicular orientation this argues against the association tracing a thin and cold cosmological filament. The timing is consistent with independent timing estimates for several suggested major galaxy encounter scenarios in the LG, during which phase-space correlated populations of TDGs could have formed that would today give rise to the observed dwarf galaxy structures \citep{2012MNRAS.423.1109P,2013MNRAS.431.3543H,2013A&A...557L...3Z}.
      \item The association aligns with the other three distant ($> 300$\,kpc) LG galaxies in the northern hemisphere of the MW in a narrow plane (RMS height of $\Delta\,=\,53$\,kpc and axis ratios of $c/a\,=\,0.10$\ and $b/a\,=\,0.76$). This ``Great Northern Plane'' passes through the center of the LG, is inclined to LGP1$^{\mathrm{mod}}$\ by only $27^\circ$ and to the GPoA by $40^\circ$ and is consistent with being aligned with the preferred orbital plane of the MW satellites in the VPOS.
      \item Five out of seven (6 of 8 if the later discovered Leo P would be included) of the galaxies in the Great Northern Plane have been identified as likely backsplash objects by \citet{2012MNRAS.426.1808T}, and most of the remaining five backsplash candidates in the southern Galactic hemisphere are also situated close to the same plane. As only a small fraction of sub-haloes in simulations are identified as backsplash objects the finding of a majority of such galaxies in one hemisphere is extremely unlikely ($\approx 0.1$\ per cent) and might constitute an ``overabundant backsplash problem'' for $\Lambda$CDM. It would mean that more galaxies are receding in one direction from the MW than are being accreted onto the MW from that direction. A natural explanation for this would be the local formation of galaxies as TDGs, which, if expelled to large distances, have very similar orbital properties like cosmological backsplash galaxies. That the backsplash candidates preferentially lie in a common plane is also consistent with an interpretation as phase-space correlated TDGs.
      \item LG galaxies are found to be preferentially infalling in the Galactic south and receding in the Galactic north, which possibly indicates that the MW is moving through a stream of dwarf galaxies. This would be in qualitative agreement with the M31-merger scenario by \citet{2010ApJ...725..542H}, in which our Galaxy passes through the tidal debris expelled in a past merger forming M31 \citep[see also][]{2010ApJ...725L..24Y,2012MNRAS.427.1769F,2013MNRAS.431.3543H}. However, the TDG origin would be in M31, such that the galaxies move away from M31 sufficiently fast which would imply that they have a tangential velocity component relative to the MW, rendering the timing estimate of Sect. \ref{sect:timing} somewhat useless. The scenario should however be testable on LG scales if proper motion measurements of the distant dwarf galaxies could be obtained. 
   \end{enumerate}

\section*{Acknowledgements}
We thank Heather Morrison and Paul Harding for very useful discussions and Shea Garrison-Kimmel for making the ELVIS simulations publicly available. We also thank the referee, Rodrigo Ibata, for very helpful comments.

\footnotesize{
\bibliographystyle{mn2e}
\bibliography{references}
}

\label{lastpage}

\end{document}